\documentclass[pra,twocolumn,showpacs,preprintnumbers,amsmath,amssymb]{revtex4}
\usepackage{mathrsfs}
\usepackage{graphicx}

\begin{document}

\title{Thermodynamics of rotating Bose gases in a trap}

\author{Jinghan Fan$^{1,2}$}
\author{Qiang Gu$^{2,3}$}\email[Corresponding author: ]{qgu@ustb.edu.cn}
\author{Wei Guo$^{1}$}
\affiliation{ $^1$School of Physics, Peking University,  Beijing 100871, China\\
$^2$Department of Physics, University of Science and Technology
Beijing, Beijing 100083, China\\
$^3$Kavli Institute for Theoretical Physics China, CAS, Beijing
100190, China }

\date{\today}

\begin{abstract}

Novel ground state properties of rotating Bose gases have been
intensively studied in the context of neutral cold atoms. We
investigate the rotating Bose gas in a trap from a thermodynamic
perspective, taking the charged ideal Bose gas in magnetic field
(which is equivalent to a neutral gas in a synthetic magnetic field)
as an example. It is indicated that the Bose-Einstein condensation
temperature is irrelevant to the magnetic field, conflicting with
established intuition that the critical temperature decreases with
the field increasing. The specific heat and Landau diamagnetization
also exhibit intriguing behaviors. In contrast, we demonstrate that
the condensation temperature for neutral Bose gases in a rotating
frame drops to zero in the fast rotation limit, signaling a
non-condensed quantum phase in the ground state.

\end{abstract}

\pacs{03.75.Hh, 03.75.Lm, 05.30.Jp, 75.20.-g}

\maketitle

The investigation of rotating quantum gases is one of the central
topics in the study of superfluidity and superconductivity. So far,
research attention in this regard mainly focuses on the physics of
quantized vortices in superfluids or superconducting samples.
Especially in recent years, several experiments using rotating
Bose-Einstein condensates (BEcs) of trapped alkali atoms have
provided spectacular illustrations of the notion of quantized
vortices \cite{Madison2000, Abo-Shaeer2001, Haljan2001, Hodby2001,
Bretin2004, Schweikhard2004}, and thus stimulate enormous interest
in the properties of rotating condensates \cite{Fetter2009}. In
these experiments, the atomic gas is equivalent to being confined in
a rotating frame, where the Hamiltonian of a single particle is
given by
\begin{align}\label{Hxy1}
H_{xy} &= \frac {P^2}{2M} + \frac 12 {M\omega_0^2 r^2} - \Omega L_z
\nonumber\\
  &= \frac{1}{2M}(\boldsymbol{P} - \boldsymbol{A})^2
    +\frac{1}{2}M(\omega^2_0-\Omega^2) r^2,
\end{align}
with $\omega_0$ denoting the frequency of harmonic potential in the
x,y plane and $r^2=x^2+y^2$. $\Omega$ is the rotational angular
frequency around the z axis, which results in an effective vector
gauge potential, $\boldsymbol{A}=(-M\Omega y,M\Omega x,0)$. As the
rotation frequency $\Omega$ increases, one or several vortices are
formed in the condensate. With more rapid rotation, the Abrikosov
vortex lattice can be observed \cite{Abo-Shaeer2001, Bretin2004,
Schweikhard2004}.

According to the Hamiltonian (\ref{Hxy1}), the regime where $\Omega$
near the trap frequency $\omega_0$ (called the fast-rotation regime)
is of special interest \cite{Fetter2009, Bloch2008}. As
$\Omega\to\omega_0$, the single-particle energy levels become
macroscopically degenerate and the rotating dilute gas is expected
to exhibit novel quantum phases analogous to the quantum Hall state
of electrons in the strong perpendicular magnetic field, if
interactions between atoms are taken into account \cite{Ho2001,
Fischer2003, Watanabe2004, Wilkin2000, Cooper2001}. During this
process, a quantum phase transition from the superfluid state
(Bose-condensed state at the mean-field quantum Hall regime)
\cite{Ho2001, Fischer2003, Watanabe2004} to the highly correlated
non-condensate state (referred as the fractional quantum Hall
regime) \cite{Wilkin2000, Cooper2001} takes place.

An alternative approach to spin up neutral atomic gases is
introducing a ``synthetic" magnetic field \cite{Higbie2002,
Jaksch2003, Juzeliunas2004}, instead of rotating the frame. This
approach creates effective gauge potentials $\boldsymbol{A}$ for
atoms by dressing them in a spatially dependent manner with optical
field that couples different atomic internal states. The effective
Hamiltonian for a trapped neutral atom in the ``synthetic" field is
like that of charged particles in the magnetic field,
\begin{eqnarray}\label{Hxy2}
H_{xy} = \frac{1}{2M}(\boldsymbol{P} - \boldsymbol{A})^2
    + \frac{1}{2}M\omega_0^2 (x^2+y^2) + V_{\rm extra}(\boldsymbol{r}) .
\end{eqnarray}
Here $V_{\rm extra}$ is the modification to the trapping potential,
which is related to the optical field where $\boldsymbol{A}$ is
induced \cite{Juzeliunas2004}. Both $\boldsymbol{A}$ and $V_{\rm
extra}$ can be engineered flexibly through constructing the
position-dependence of the dressed atomic states
\cite{Juzeliunas2004, Lin2009a}. Quite recently, the light-induced
magnetic field is realized in experiments, with producing either
uniform vector potential \cite{Lin2009a} or spatially-varied one
\cite{Lin2009b}. The latter corresponds to a nonzero magnetic field
which can generate vortices in the ground state of BEcs, and this
phenomenon has just been indeed observed \cite{Lin2009b}. An
advantage of this optical approach over rotating gases is that the
synthetic field can be significantly large, making possible the
study of the fractional quantum Hall regime.

Both approaches mentioned above acquires a vector potential for
neutral atoms, so as to imitate charged particles in the magnetic
field. Actually, the charged Bose gas (CBG) has already been studied
for more than half a century. Earlier in 1950s, it showed that the
ideal CBG exhibits essential equilibrium features of a
superconductor \cite{Blatt1955, Schafroth1955}. Furthermore, it was
indicated that an arbitrarily small value of the magnetic field can
eliminate Bose-Einstein condensation (BEC) in a 3-dimensional ideal
CBG, while the orbital motion results in extremely large Landau
diamagnetism and even leads to Meissner effect at low temperatures
\cite{Schafroth1955, Else, Alexandrov, Daicic, Toms}.

In this paper, we study properties of rotating Bose gases in a trap
from a thermodynamic perspective. Without loss of generality, we
consider a ideal CBG trapped in the harmonic potential as an
example. The obtained results are applicable for neutral Bose gases
either in a rotating frame or in a synthetic magnetic field. We
demonstrate that these two cases display distinct thermodynamic
properties.

 {\sl The ideal Bose gas model.} --
Let us consider an ideal Bose gas with mass $M$ and charge $q$. The
whole system is placed in a uniform magnetic field
$\boldsymbol{B}=B\hat{\boldsymbol{e}}_z$ and an anisotropic harmonic
potential. Choosing the gauge $ \boldsymbol{A} =
\frac{1}{2}\boldsymbol{B}\times \boldsymbol{r}$ for the vector
potential, the Hamiltonian can be expressed as $H=H_z+H_{xy}$, where
the Hamiltonian
\begin{eqnarray}\label{Hz}
H_z= \frac{P_z^2}{2M}+\frac{1}{2}M\omega_z^2z^2
\end{eqnarray}
describes the z-direction with trap frequency $\omega_z$ and
\begin{eqnarray}\label{Hxy3}
H_{xy}=
\frac{\hbar^2}{2M}(P_x^2+P_y^2)+\frac{1}{2}M\omega^2(x^2+y^2)-\omega_ll_z
\end{eqnarray}
describes the x,y plane, where $\omega=\sqrt{\omega_l^2+\omega_0^2}$
with $\omega_l=\frac{qB}{2Mc}$. Equation (\ref{Hxy3}) can be derived
from Eq. (\ref{Hxy2}) without the $V_{\rm extra}$ term. Thus the
eigenvalues of quantized levels for a boson are of the form
\begin{align}\label{epsilon}
\bar{\epsilon}_{n_z,n_\rho,m} = n_z+\frac{1}{2} +
  \left( 2n_\rho+|m|+1\right) \sqrt{\alpha^2+\bar{B}^2}
    - m\bar{B}
\end{align}
with $n_z=0,1,2,...$, $n_\rho=0,1,2,...$, and $m=0$, $\pm 1$, $\pm
2,...$, where some dimensionless variables are introduced:
$\bar{\epsilon}_{n_z,n_\rho,m}={\epsilon_{n_z,n_\rho,m}}/{(\hbar\omega_z)}$,
$\bar{B}={\omega_l}/{\omega_z}$ and $\alpha={\omega_0}/{\omega_z}$.
Note that $\bar{B}$ is proportional to the external magnetic field
(synthetic magnetic field) for charged (neutral) particles.
Generally there is no degeneration of energy levels, except that the
accidental degeneracy arises when at least one of $\bar{B}$,
$\sqrt{\alpha^2+\bar{B}^2}$ and
$\bar{B}/{\sqrt{\alpha^2+\bar{B}^2}}$ is a rational number.
$\bar{\epsilon}_{0} = \bar{\epsilon}_{0,0,0} =
\frac{1}{2}+\sqrt{\alpha^2+\bar{B}^2}$ is the ground state energy of
one boson.

Now we consider an assembly of $N$ bosons, whose thermodynamic
potential is written as $\bar{\Omega} = \bar{\Omega}_{\rm T} +
\bar{\Omega}_0$, where
\begin{eqnarray}
\bar{\Omega}_{\rm T} =
\bar{T}{\sum_{n_z,n_\rho,m}}\ln\left(1-e^{-\frac{\bar{\epsilon}_{n_z,n_\rho,m}-\bar{\mu}}{\bar{T}}}\right)
\end{eqnarray}
with the dimensionless temperature $\bar{T}=k_BT/(\hbar\omega_z)$
and chemical potential $\bar{\mu}=\mu/(\hbar\omega_z)$, and
$\bar{\Omega}_0$ is the thermodynamic potentials for condensed
bosons. $\bar{\Omega}_0$ is present only when there is a nonzero
condensate described by a background field $\bar{\Psi}$, which reads
\begin{eqnarray}\label{Omega0}
\bar{\Omega}_0=\frac{1}{\hbar\omega_z}\int d^3r
   \left\{\frac{|\boldsymbol{D}\bar{\Psi}|^2}{2M}+V(\boldsymbol{r})|\bar{\Psi}|^2
     -\mu|\bar{\Psi}|^2\right\}~.
\end{eqnarray}
Here $\boldsymbol{D}\bar{\Psi} = \hbar\nabla \bar{\Psi} +
i\frac{q}{c}\boldsymbol{A}\bar{\Psi}$ is the gauge-covariant
derivative and $V(\boldsymbol{r})$ is the trapping potential as in
Eqs. (\ref{Hxy2}) and (\ref{Hz}). The presence of a condensate is
signalled by a nonzero value for $\bar{\Psi}$ which satisfies
\begin{eqnarray}\label{Schro1}
\frac{\delta\bar{\Omega}}{\delta{\bar{\Psi}}^*} = \frac{1}{2M}
    \boldsymbol{D}^2\bar{\Psi}-V(\boldsymbol{r})\bar{\Psi}+\mu\bar{\Psi}=0.
\end{eqnarray}
$\bar{\Psi}$ can be expanded as $\bar{\Psi}(\boldsymbol{r})=\sum_n
C_n f_n(\boldsymbol{r})$, where $f_n(\boldsymbol{r})$ are stationary
state solutions to the Schr\"odinger equation,
\begin{eqnarray}\label{Schro2}
-\frac{1}{2M}\boldsymbol{D}^2f_n(\boldsymbol{r})+V(\boldsymbol{r})f_n(\boldsymbol{r})
=E_nf_n(\boldsymbol{r}).
\end{eqnarray}
and they form a complete set. For a ideal gas, the eigenvalues $E_n$
coincide with the one particle energy described by Eq.
(\ref{epsilon}). Combining Eqs. (\ref{Schro1}) and (\ref{Schro2}),
the coefficients $C_n$ are determined by $(E_n-\mu)C_n=0$. The
chemical potential is smaller than the ground state energy,
$\mu<\epsilon_0$, above the BEC temperature $\bar{T}_c$, and
therefore $E_n-\mu\ne 0$ and $C_n=0$ for all levels. With the
temperature decreasing, $\mu$ approaches $\epsilon_0$ until
$\mu=\epsilon_0$ at or below $\bar{T}_c$. Then the only solution
$C_{n=0}\ne 0$ exists at $\bar{T}<\bar{T}_c$ and the condensate wave
function is nonzero correspondingly, $\bar{\Psi}(\boldsymbol{r})=C_0
f_0(\boldsymbol{r})$, where
$f_0=b\mathscr{F}(0,1,a_l^2\rho^2)e^{-\frac{1}{2}(a_l^2\rho^2+a_z^2z^2)}$
is the normalized eigenfunction corresponding to the lowest energy
$\epsilon_0$. $\mathscr{F}(\alpha, \gamma, x)=\sum_{k=0}^\infty
\frac{1}{k!}\frac{\alpha(\alpha+1)...(\alpha+k-1)}{\gamma(\gamma+1)...(\gamma+k-1)}x^k$
is confluent hypergeometric function. $b$ represents the normalizing
constant. $a_l=\sqrt{{M|\omega_l|}/{\hbar}}$ and
$a_z=\sqrt{{M\omega_z}/{\hbar}}$.

 {\sl The BEC temperature and condensate fraction.} --
The particle number is derived from the thermodynamic potential by
the standard procedure
$N=-{\partial\bar{\Omega}}/{\partial\bar{\mu}}$,
\begin{equation}\label{number}
N=
\begin{cases}
\sum \left( \exp{\frac{\bar{\epsilon}_{n_z,n_\rho,m}
  -\bar{\mu}}{\bar{T}}}-1 \right)^{-1}&\text{above } \bar{T}_c~; \\
\sum^\prime \left(\exp{\frac{\bar{\epsilon}_{n_z,n_\rho,m}
  -\bar{\mu}}{\bar{T}}}-1 \right)^{-1}+N_0&\text{below } \bar{T}_c~.
\end{cases}
\end{equation}
Here $\sum$ denotes the summation over all the eigen states but the
ground state is not included in $\sum^\prime$. $N_0$ is the number
of particles condensed on the ground state,
\begin{eqnarray}
N_0 = -\frac{\partial\bar{\Omega}_0}{\partial\bar{\mu}}=|C_0|^2.
\end{eqnarray}
Formally, the dimensionless chemical potential $\bar{\mu}$ is a
function of the external magnetic field $\bar{B}$ and the
temperature $\bar{T}$, which can be obtained from Eq.
(\ref{number}). As discussed above,
$\bar{\epsilon}_0-\bar{\mu}\rightarrow0$ as
${\bar{T}}\rightarrow{\bar{T}_c^+}$. Above $\bar{T}_c$, we obtain an
analytical expression for the particle number by converting the
summation in Eq. (\ref{number}) into definite integral
\cite{Pethick},
\begin{eqnarray}\label{number-c}
N &=& \frac{\bar{T}^3g_3(\frac{\bar{\epsilon}_0-\bar{\mu}}{\bar{T}})}
   {2\sqrt{\alpha^2+\bar{B}^2}} \left(\frac1{\sqrt{\alpha^2+\bar{B}^2}+\bar{B}}
     + \frac1{\sqrt{\alpha^2+\bar{B}^2}-\bar{B}} \right) \nonumber\\
  &=& \frac{\bar{T}^3}{\alpha^2}g_3(\frac{\bar{\epsilon}_0-\bar{\mu}}{\bar{T}})~,
\end{eqnarray}
where $g_\gamma (z)$ is the polylogarithm function defined as
$g_\gamma (z)=\sum_{n=1}^\infty \frac{e^{-nz}}{n^\gamma}$. The
series converge when $\gamma>1$ and $z>0$. $g_\gamma
(0)=\zeta(\gamma)$ is just the Riemann zeta function. The BEC
temperature $\bar{T}_c$ is determined by setting
$\bar{\epsilon}_0-\bar{\mu}$ to be zero in Eq. (\ref{number-c}),
that is $N={\bar{T}_c^3}g_3(0)/{\alpha^2}$ where $g_3(0)=1.202$. So
we have
\begin{eqnarray}\label{Tc}
\bar{T}_c = \left(\frac{\alpha^2N}{g_3(0)}\right)^\frac{1}{3}~.
\end{eqnarray}
This is just the BEC temperature for a Bose gas of $N$ particles
confined in an anisotropic trap with the frequencies $\omega_0$ in
x,y plane and $\omega_z$ in the z-direction. Since the one-particle
energy spectra has already significantly changed by the external
magnetic field, as suggested in Eq. (\ref{epsilon}), it is very
surprising that {\it the BEC temperature $\bar{T}_c$ is irrelevant
to the magnetic field $\bar{B}$}. This result seems conflict with
our intuition established in the study of superconductivity, where
the transition temperature decreases as the magnetic field is
strengthened. Below $\bar{T}_c$, the number of thermal particles is
$N_T = \frac{\bar{T}^3}{\alpha^2} g_3(0) =
N(\frac{\bar{T}}{\bar{T_c}})^3$, so the condensate fraction reads
\begin{eqnarray}
n_0 = \frac{N_0}N = 1-\frac{N_{exc}}N
    = 1-\left(\frac{\bar{T}}{\bar{T_c}}\right)^3~.
\end{eqnarray}
Again, $n_0$ varies as if the magnetic field does not exist.

We now consider a Bose gas confined in a rotating frame, as
described by Eqs. (\ref{Hxy1}) and (\ref{Hz}). Similarly, we obtain
the BEC temperature,
\begin{eqnarray}\label{Tc}
\bar{T}_c=\left(\frac{\omega^2_0-\Omega^2}{\omega^2_zg_3(0)}N\right)^\frac{1}{3}~.
\end{eqnarray}
With the rotation frequency $\Omega$ rising up to $\omega_0$, the
BEC temperature goes down to zero. This result is consistent with
the predication that the ground state in the fast rotation regime
undergoes a quantum transition to a non-condensed state as
$\Omega\to\omega_0$  \cite{Wilkin2000, Cooper2001}.

It is worth noting that Eq. (\ref{Tc}) is also helpful to understand
the magnetic properties of CBGs. With $\Omega$ approaching
$\omega_0$, the trapping is canceled by the centrifugal effect and
the rotating Bose gas reduces to a homogenous CBG in a constant
external field, as Eq. (\ref{Hxy1}) indicates. It is well
established that the BEC does no longer take place in this case
\cite{Else, Alexandrov, Daicic, Toms}.

 {\sl The internal energy and specific heat.} --The internal energy is
calculated for two cases separately. For Case I that
$\bar{T}>\bar{T}_c$, the internal energy is given by
\begin{eqnarray}\label{UT}
\bar{U}&=& \bar{\epsilon}_0 N + \sum \frac{\bar{\epsilon}_{n_z,n_\rho,m}-\bar{\epsilon}_0}
    {\exp\frac{{\bar{\epsilon}_{n_z,n_\rho,m}-\bar{\mu}}}{\bar{T}}-1} \nonumber\\
  &=& \bar{\epsilon}_0 N
     + \frac{3\bar{T}^4}{\alpha^2}g_4(\frac{\bar{\epsilon}_0-\bar{\mu}}{\bar{T}})~.
\end{eqnarray}
For Case II that $\bar{T}\le \bar{T}_c$, $\bar{U}$ is expressed as
\begin{eqnarray}\label{UT0}
\bar{U} &=& \bar{\epsilon}_0 N_0 + \bar{\epsilon}_0 N_{\rm T} +
   {\sum}^\prime \frac{\bar{\epsilon}_{n_z,n_\rho,m}-\bar{\epsilon}_0}
   {\exp\frac{{\bar{\epsilon}_{n_z,n_\rho,m}-\bar{\epsilon}_0}}{\bar{T}}-1} \nonumber\\
 &=& \bar{\epsilon}_0 N + \frac{3\bar{T}^4}{\alpha^2}g_4(0).
\end{eqnarray}
Correspondingly, the specific heat can be derived from the internal
energy,
\begin{equation}\label{specific}
\bar{C}=\begin{cases}
12N\frac{g_4(\frac{\bar{\epsilon}_0-\bar{\mu}}{\bar{T}})}{g_3(0)}(\frac{\bar{T}}{\bar{T}_c})^3
   - 9N\frac{g_3(\frac{\bar{\epsilon}_0-\bar{\mu}}{\bar{T}})}
     {g_2(\frac{\bar{\epsilon}_0-\bar{\mu}}{\bar{T}})} &\text{above } \bar{T}_c~; \\
12N\frac{g_4(0)}{g_3(0)}(\frac{\bar{T}}{\bar{T}_c})^3~ &\text{below
} \bar{T}_c~.
\end{cases}
\end{equation}

\begin {figure}[t]
\includegraphics[width=0.4\textwidth,keepaspectratio=true]{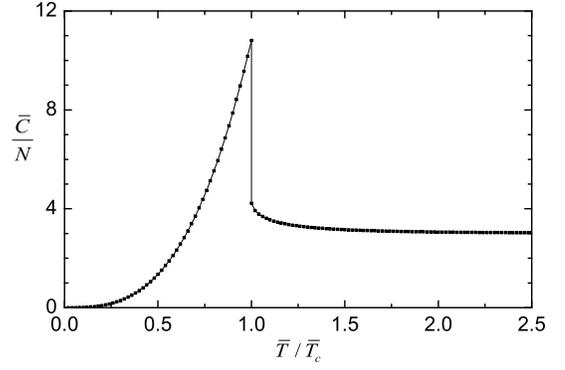}
\caption{The specific heat $\bar{C}/N$ as a function of normalized
temperature $\bar{T}/\bar{T}_c$ for charged Bose gases in magnetic
field. This curve is universal to all the trapped Bose
gases.}\label{CT}
\end{figure}

At $\bar{T}>\bar{T}_c$, $\bar{\epsilon}_0-\bar{\mu}$ satisfies Eq.
(\ref{number-c}) and it is just the function of temperature
$\bar{T}$, while $\bar{\epsilon}_0-\bar{\mu}=0$ at
$\bar{T}\le\bar{T}_c$. In both cases, $\bar{\epsilon}_0-\bar{\mu}$
is independent on the magnetic field $\bar{B}$. So, the
$\bar{B}$-dependency of $\bar{U}$ in Eqs. (\ref{UT}) and (\ref{UT0})
is just due to the $\bar{\epsilon}_0N$ term where $\bar{\epsilon}_0$
is in relation to $\bar{B}$. Nevertheless, since the
$\bar{\epsilon}_0N$ term remains constant as the temperature varies,
the specific heat has no relation with $\bar{B}$. Figure \ref{CT}
shows the specific heat as a function of the normalized temperature,
which displays a discontinuity at the BEC temperature.

 {\sl The Landau diamagnetism.} --
We now calculate the magnetization according to
$\bar{M}=-{\partial\bar{\Omega}}/{\partial\bar{B}}$, which yields
\begin{eqnarray}\label{MT}
\bar{M} &=& -\sum \frac{
\frac{(2n_\rho+|m|)\bar{B}}{\sqrt{\alpha^2+\bar{B}^2}}-m}
     {\exp\frac{{\bar{\epsilon}_{n_z,n_\rho,m} - \bar{\mu}}}{\bar{T}}-1}
    -\frac{N\bar{B}}{\sqrt{\alpha^2+\bar{B}^2}} \nonumber\\
   &=& - \frac{N\bar{B}}{\sqrt{\alpha^2+\bar{B}^2}}
\end{eqnarray}
at $\bar{T}>\bar{T}_c$. Below $\bar{T}_c$, $\bar{M}$ consists of two
parts. One is due to the condensed particles,
\begin{eqnarray}
\bar{M}_0 = -\frac{\partial\bar{\Omega}_0}{\partial \bar{B}}
    =-\int d^3r \frac{\partial\bar{\epsilon}_0}{\partial \bar{B}}|\bar{\Psi}|^2
    =-\frac{N_0\bar{B}}{\sqrt{\alpha^2+\bar{B}^2}}~.
\end{eqnarray}
The other is the contribution from thermal particles,
\begin{eqnarray}
\bar{M}_{\rm T} = -\frac{\partial\bar{\Omega}_{\rm T}} {\partial\bar{B}}
   = - \frac{N_{\rm T}\bar{B}}{\sqrt{\alpha^2+\bar{B}^2}}.
\end{eqnarray}
Note that the summation of both parts,
\begin{eqnarray}\label{MT0}
\bar{M} = \bar{M}_0+\bar{M}_{\rm T}
        = -\frac{N\bar{B}}{\sqrt{\alpha^2+\bar{B}^2}}~,
\end{eqnarray}
just amounts to the result above $\bar{T}_c$. In addition, the
susceptibility has the following form,
\begin{eqnarray}\label{suscept}
\bar{\chi}=\frac{-N\alpha^2}{(\alpha^2+\bar{B}^2)^\frac{3}{2}}~.
\end{eqnarray}
It needs points out that {\it both the magnetization and
susceptibility keep invariant at all temperatures}. According to
Eqs. (\ref{MT}), (\ref{MT0}) and (\ref{suscept}), the magnetization
and susceptibility are negative, reflecting that the gas exhibits
the {\it Landau diamagnetism}.

The diamagnetization for a homogeneous CBG in the magnetic field has
been investigated by several groups \cite{Alexandrov, Daicic, Toms}.
The magnetization $\bar{M}$ varies with the temperature and the
diamagnetism is stronger at lower temperatures. Especially, below
the BEC temperature, $\bar{M}$ does not vanish as the external field
$\bar{B}$ is reduced to zero, which is the evidence of the
Meissner-Ochsenfeld effect. Nevertheless, present results shows that
$\bar{M}$ vanishes as $\bar{B}\to 0$ at all temperatures, implying
that the Meissner-Ochsenfeld effect might not exist in a trapped
CBG. The trapping potential brings about significant changes to the
physics of the charged Bose gases.

In summary, we show that the charged Bose gas confined in a harmonic
trap undergoes Bose-Einstein condensation at a critical temperature
determined by the trapping potentials and irrelevant to the magnetic
field, although the application of the field changes the energy
spectrum of bosons. The specific heat is also independent of the
external field. Moreover, we find that the Landau diamagnetization
is strengthened with the magnetic field, but keeps unchanging at all
temperatures. Our results are applicable to the neutral atomic Bose
gas which is caused to rotate either by a rotating frame or due to a
synthetic magnetic field. The latter behaves like a charged Bose gas
in magnetic field, while the Bose gas in a rotating frame displays
distinct thermodynamic behaviors. Its Bose-Einstein condensation
temperature does no longer hold a constant but decreases with the
rotating frequency increasing. The condensation does not occur once
the rotation frequency approaches the trap frequency, consistent
with the predication that the ground state in the fast rotation
regime is a non-condensed state.

At last, we briefly discuss the possibility to realize charged Bose
gases in experiments. It is already possible to create ultracold
plasmas by photoionization of laser-cooled neutron atoms
\cite{Killian1999}. The temperatures of electrons and ions are as
low as 100 mK and 10 $\mu$K, respectively. The ions can be regarded
as charged bosons if their spin is an integer.

This work was supported by the Fok Ying Tung Education Foundation of
China (No.~101008), the Key Project of the Chinese Ministry of
Education (No.~109011), and the Fundamental Research Funds for the
Central Universities of China.

\end{document}